\documentclass[aps,pre,onecolumn]{revtex4}
\usepackage{graphicx}
\usepackage{amsmath,amsfonts,amssymb}
\usepackage{bm}
\usepackage{color}
\usepackage{soul}
\begin{document}

\title{Irreversibility and typicality: A simple analytical result for the Ehrenfest model}

\author{Marco Baldovin}
\affiliation{Dipartimento di Fisica, Universit\`a di Roma ``Sapienza'', P.le A.Moro 5 I-00185, Rome, Italy}
\author{Lorenzo Caprini}
\affiliation{Gran Sasso Science Institute (GSSI), Via F.Crispi 7, I-67100 L'Aquila, Italy}
\author{Angelo Vulpiani}
\affiliation{Dipartimento di Fisica, Universit\`a di Roma Sapienza, P.le Aldo Moro 5, 00185, Rome, Italy}
\affiliation{Centro Linceo Interdisciplinare ``B. Segre'', Accademia dei Lincei, Rome, Italy}

\date{\today}

\begin{abstract}
With the aid of simple analytical computations for the Ehrenfest model, we  
clarify some basic features of macroscopic irreversibility. The 
stochastic character of the model allows us to give a non-ambiguous 
interpretation of the general idea that irreversibility is a typical property: 
for the vast majority of the realizations of the stochastic process, a single trajectory of  a 
macroscopic observable behaves irreversibly, remaining ``very close'' to the 
deterministic evolution of its ensemble average,  which can be computed using 
probability theory. The validity of the above scenario is checked through simple 
numerical simulations and  a rigorous proof  of the typicality is provided in 
the thermodynamic limit.
\end{abstract}

\maketitle

\section{Introduction}

Understanding the irreversibility from first principles is an old and noble 
problem of Physics. The technical reason of its difficulty is rather clear: on 
the one hand, the microscopic world is ruled by laws (Hamilton equations) which 
are invariant under the transformation of time reversal ($t \to - t \,\, , \,\, 
{\bf q} \to  {\bf q} \,\, , \,\, {\bf p} \to - {\bf p}$, being ${\bf q}$ and 
${\bf p}$ the positions and momenta of the system); the macroscopic world, on 
the other hand, is described by irreversible equations, e.g. the Fick equation 
for the diffusion~\cite{boltzmann03, cercignani98, emch02, castiglione08}. 

How is it possible to conciliate the two above facts? On this topic there  is an 
aged debate which started with the celebrated Boltzmann's $H$ theorem and the 
well known criticisms by Loschmidt (about reversibility) and Zermelo (about 
recurrency). Of course we cannot  enter in a detailed discussion about this 
fascinating chapter of statistical mechanics~\cite{cercignani98, zanghi05}. Already Boltzmann and  
Smoluchowski understood that the criticism by Zermelo is not a real serious 
problem as long as macroscopic systems are considered: basically, because of the 
Kac's lemma,  in macroscopic systems the recurrence time  is so large that it 
cannot be observed~\cite{cercignani98, chibbaro14, lazarovici15}. We can summarise the conclusions of Boltzmann by saying that 
the irreversibility describes an empirical regularity of macroscopic objects 
which is valid for a ``vast majority'' of the possible initial conditions. Often 
such validity for the ``vast majority'' of initial conditions is called {\it 
typicality}. According to Lebowitz~\cite{lebowitz93} (as well as many others) {\it  a certain 
behavior is typical if the set of microscopic states for which it occurs 
comprises a region whose volume fraction goes to one as the number of molecules} 
$N$ {\it grows}. We can state that irreversibility is an emergent property~\cite{berry94, chibbaro14, zanghi05}
which appears as the number of degrees of freedom 
becomes (sufficiently) large; in such a limit, a single observation of the system 
is enough to determine its macroscopic properties.

Several mathematical results, as well as detailed  numerical simulations, 
support the coherence of the scenario proposed by Boltzmann~\cite{cercignani98}. Among the others, 
Lanford's work about the rarified gases is particularly important: he 
was able to prove, in a rigorous fashion, the validity of the Boltzmann equation 
for short times (of the order of the collision time) in the so-called 
Boltzmann-Grad limit~\cite{lanford83}. 

In spite of the above mentioned results, irreversibility still remains a somehow 
misinterpreted and controversial issue. The reader may appreciate the diversity 
of  opinions from the comments~\cite{driebe94} to a well known  paper by Lebowitz on  
Boltzmann's  approach to the irreversibility. For instance, Prigogine and his 
school claim that irreversibility is either true on all levels or on none: it 
cannot emerge as if out of nothing, on going from one level to another~\cite{bricmont95, chibbaro14, driebe94}. For 
others, irreversibility either results from (microscopic) chaotic dynamics or it 
is a mere consequence of the interaction with the external environment.
 
 One source of the controversy about the Boltzmann point of view, in 
particular among philosophers of science, is how to interpret typicality~\cite{frigg09}.
  
 This article aims at supplementing, mainly for pedagogical purposes,  the basic 
aspects of Boltzmann's explanation of macroscopic irreversibility. In order to 
present a clear non ambiguous analysis we treat a stochastic  system, i.e the 
celebrated Ehrenfest model, which is nothing but a Markov chain. A  simple and 
neat  analytical computation for the  model shows, in a precise way, that for $N 
\gg 1$ each realization is very close, at any time, to the average value (which 
can be easily computed).
 
 The paper is organized as follows: in Sec.~\ref{Sec:2} we introduce the problem 
of typicality together with some remarks about ensembles and entropies.
Sec.~\ref{Sec:3} is devoted to the Ehrenfest model, and to some numerical 
results. Then, we rigorously prove the typicality of a trajectory for such a 
model in Sec.~\ref{Sec:4}. Finally we summarize the results in Sec.~\ref{Sec:5}.

\section{Remarks about ensembles, entropies and typicality}
\label{Sec:2}

Traditionally, entropy  has an important relevance in the treatment of 
irreversibility; it seems to us that this central role is mainly based on 
historical grounds. In the present paper we do not discuss irreversibility in 
terms of entropy, for two main reasons. First, the word entropy can be source of 
confusion: for instance the entropy $S_G$, defined in terms of the probability 
distribution function in the  $\Gamma$- space, has a completely different 
behavior from $S_B$, i.e the entropy obtained from the probability density of a 
single particle ($\mu$-space); for a discussion on this point see~\cite{castiglione08, cerino16, lebowitz93}. 
Second, at a practical, as well as at a conceptual level, for understanding 
of irreversibility it is enough to observe that, if the system starts from  a 
typical far-from-equilibrium initial state,  the macroscopic observables stay 
close to their mean values during the evolution, and therefore they approach 
their  equilibrium values. In Sec.~\ref{Sec:4} we will discuss this point in a 
precise mathematical way.

Even if probability theory  has a great relevance for statistical 
mechanics, it is necessary to avoid mixing mathematics and physics. It is true 
that  the building of the standard formulation of statistical mechanics is based on 
statistical ensembles, but this approach can be seen as a mere stratagem, and 
it is ultimately unconvincing in the following sense: in experiments, as well as 
in numerical computations, we are forced to treat a unique system, and we have 
not access to  a collection of identical systems~\cite{caprara18, zurek18}. At a physical level  
the  relevant  problem is: what is the  link between the probabilistic 
computations (i.e. the averages over an ensemble) and the actual results 
obtained by looking at a  single realization (or sample) of the system under 
investigation?

In particular   one should be careful to avoid the confusion between 
irreversibility and relaxation of the phase space probability distribution~\cite{cerino16, lebowitz93}. In 
presence of ``good chaotic properties'' (mixing systems) one has that the 
probability density, $\rho({\bf X},t)$,  relaxes (in a suitable technical sense) 
to the invariant distribution for large times, i.e. $\rho({\bf X},t) \to 
\rho_{inv}({\bf X})$ as $t \to \infty$. This property is remarkable, and rather 
important in the dynamical systems context; still, it cannot be considered 
physical irreversibility. Actually, from a physical point of view, the true 
question is to show that a {\it single macroscopic} system shows an irreversible 
behavior, for a ``generic'' initial state. In crude terms the interesting point 
is to understand the cooling of a single (initially hot) pot and not the 
behavior of an ensemble of pots.

In deterministic systems a  delicate   point is how to  intend  typicality i.e. 
the precise mathematical meaning of  ``vast majority''. As already mentioned, 
for many scientists active in statistical mechanics  ``vast majority'' means 
with probability close to $1$ with respect to the Lebesgue measure, in physical 
terms microcanonical distribution~\cite{goldstein12, lebowitz93, zurek18}. Such an interpretation has been considered 
not convincing by some authors who criticized the privileged status of the 
Lebesgue measure~\cite{frigg09}. Although, in our opinion, there are very good reasons to 
privilege the microcanonical distribution,  we do not insist in such 
controversial aspect. In the following we will consider a well known stochastic 
model~\cite{ehrenfest15, kac57} where there is no ambiguity about the meaning of ``vast majority''.
An analysis of simplified stochastic models of this kind can help very much, in particular at pedagogical level, for understanding irreversibility (see for instance~\cite{gottwald09} for a discussion about Kac ring model).

\section{The Ehrenfest model: heuristic results}
\label{Sec:3}

\subsection{Description and basic properties}
For the sake of self-consistency, we briefly recall the  Ehrenfest model~\cite{ehrenfest15}. We consider $N$ particles, labeled with an index $i=1,...,N$, and two 
boxes, A and B: at the beginning, each particle can be placed either in box A or 
in box B. At every time step we randomly choose an integer number between 1 and 
$N$, with a uniform distribution, and we move the corresponding particle from 
its box to the other one. The ``macroscopic'' state of the model at time $t$ is 
identified by $n_t$, the number of particles in A, while the corresponding 
``microscopic'' configuration is defined by the (labeled) particles which 
actually are in box A at that time. The Markovian evolution for the macroscopic 
state is ruled by the transition probabilities for $n_t = n$  to become $n_{t+1} 
= n \pm 1$:
\begin{equation}
\label{eq:markov}
\begin{cases}
 P_{n \to n-1}= {n \over N} \\
P_{n \to n+1}= 1-{n \over N} \,.
\end{cases}
\end{equation}
As a consequence, for any starting value $n_0$, during $t$ steps the macroscopic 
state can realize $2^t$ different trajectories. Let us notice that in the 
Ehrenfest model the detailed balance holds: this property in the stochastic 
context is somehow equivalent to the time reversibility.

\begin{figure}[!h]
\centering
\includegraphics[clip=true,width=0.7\columnwidth,keepaspectratio]{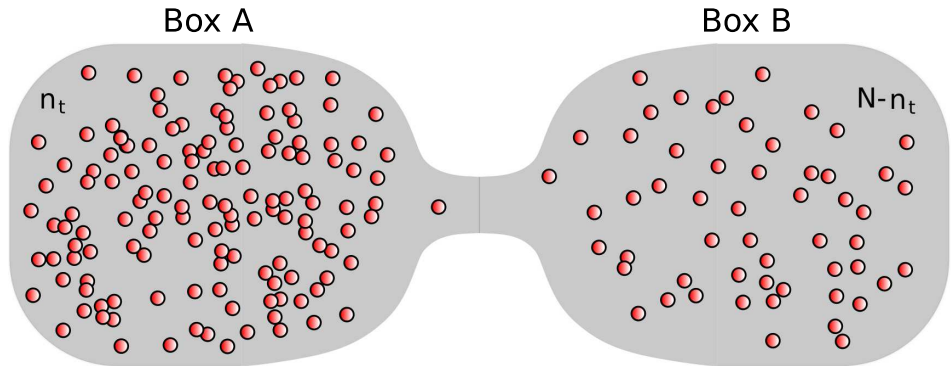}
\caption{Physical interpretation of the Ehrenfest model:  box A (containing $n_t$ particles) is connected to the box B (containing $N-n_t$ particles) by a narrow channel.}
\label{fig:sketch}
\end{figure}

This model can be  seen as a crude description of a system with $N$ particles in 
two vessels (A and B), connected by a narrow pipe, as shown in 
Fig.~\ref{fig:sketch}. Of course in this case the true dynamics is 
deterministic, while in the Ehrenfest model the evolution is stochastic.  As a 
link between the original, Hamiltonian system and its Markovian counterpart, we 
can imagine to associate to each realization of the Ehrenfest model a set of 
initial conditions in the deterministic system.

The simplicity of the model allows us to study the statistical features of the 
evolution of an ensemble of (microscopic) initial conditions in the same 
(macroscopic) state, $n_0$, by computing the evolution of the first and the 
second conditional momenta of the state $n_t$, namely  $\langle n_t\rangle$ and 
$\sigma_t^2=\langle n_t^2\rangle- \langle n_t\rangle^2$; for the sake of 
simplicity, we omit the conditional argument in the average $\langle \cdot | n_0 
\rangle$. It is easy, see \ref{Appendix:1}, to show that:
\begin{flalign}
\label{eq:conditionaln}
\langle n_t\rangle &= {N \over 2} + \left(1- {2 \over N} \right)^t\Delta_0 \,\, , \\
\label{eq:conditionalvar}
\sigma_t^2 &={N \over 4} + \left(1- {4 \over N} \right)^t\left(\Delta_0^2 - {N \over 4}\right)+
\left(1- {2 \over N} \right)^{2t} \Delta_0^2 \,,
\end{flalign}
where $\Delta_0=n_0 - N/2$.
From Eq.\eqref{eq:conditionaln} is clear that  $\langle n_t \rangle$  relaxes 
monotonically to the equilibrium (mean) value $n_{eq}=N/2$, with an exponential 
decay ruled by the characteristic time $\tau_c= - 1/\ln(1- 2/N)\simeq N/2$. In a 
similar way the conditional standard deviation $\sigma_t$ tends to its 
equilibrium value $\sqrt{N}/2$ with a characteristic time $O(N)$.

\subsection{First numerical clues of typicality}

Let us note that Eqs.\eqref{eq:conditionaln} and 
\eqref{eq:conditionalvar} provide a description of the process 
only at an average level, giving no information about the single realization. We 
will see that under the assumption $N\gg 1$ (somehow 
equivalent to the \textit{thermodynamic limit} in real physical systems) almost 
all actual realizations are arbitrarily close to the average evolution at any time, i.e. 
almost all trajectories are ``typical''.

\begin{figure}[!h]
\centering
\includegraphics[clip=true,width=0.9\columnwidth,keepaspectratio]{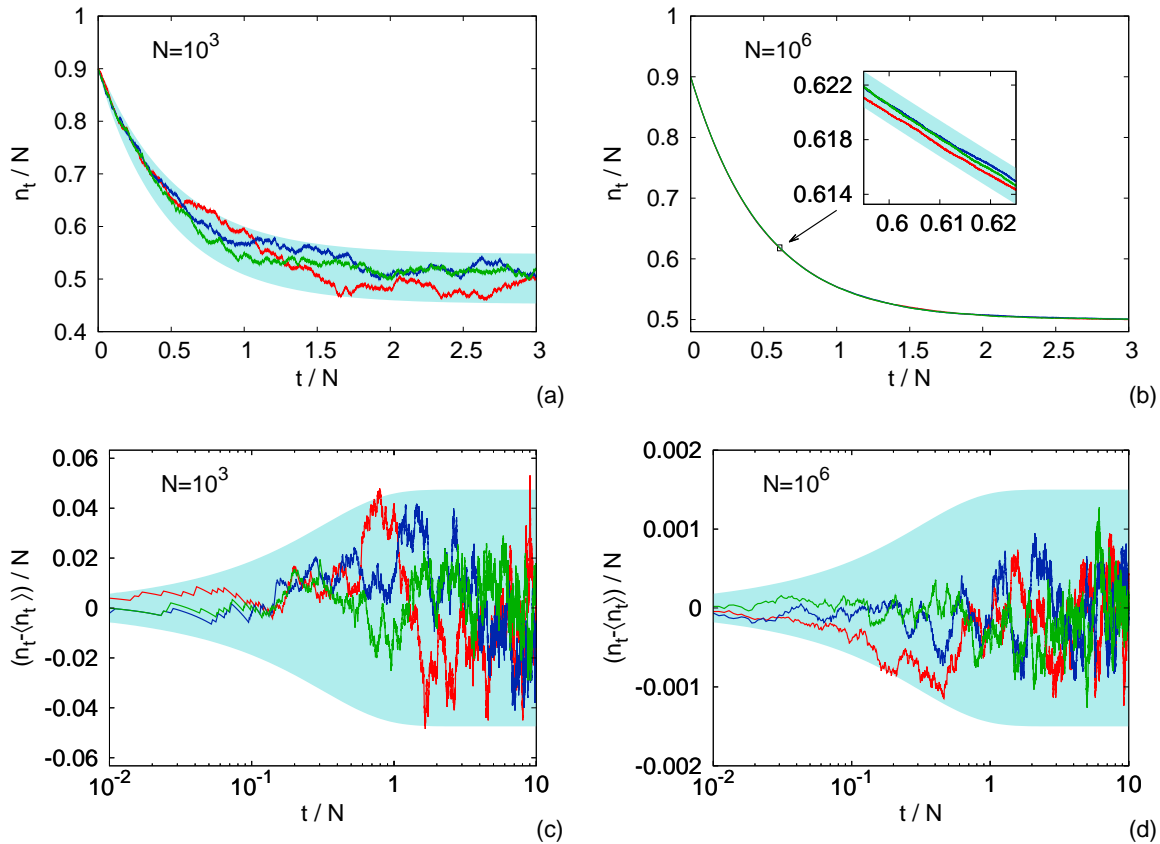}
\caption{ Single trajectories of the model. Above: $n_t/N$ as a function of the 
time $t$, for $N=10^3$ (panel (a)) and $N=10^6$ (panel (b)); three different 
realizations are shown for each case.  The inset in panel (b) is a magnified 
detail of the plot. Below: $(n_t - \langle n_t\rangle)/N$ vs $t$ for the same 
trajectories ($N=10^3$ in panel (c) and $N=10^6$ in panel (d)). Light-blue 
regions are stripes of size $\pm 3\sigma_t$ around the mean value $\langle n_t 
\rangle$. In all considered cases, $n_0=0.9N$.}\label{fig:traj}
\end{figure}

 Simple numerical computations suggest that a single trajectory is  typical in 
the sense discussed above, i.e. that behaviors very different from the average 
one are extremely rare. In Fig.\ref{fig:traj} we show $n_t/N$ in function of 
time, panels (a) and (b), for several single realizations of $n_t$ for the same 
$n_0$, where the exponential behavior clearly emerges for each trajectory. 
Panels (c) and (d), on the other hand, display $n_t - \langle n_t \rangle$ vs 
$t$ for the same trajectories; we superimpose a confidence interval (light blue 
region) obtained by considering a stripe~$\pm 3\sigma_t$ around the conditional 
average value, $\langle n_t\rangle$, as computed in Eqs.~\eqref{eq:conditionaln} 
and \eqref{eq:conditionalvar}. Each trajectory is contained in this stripe for 
almost all times: since the trajectories are closer to their mean value as $N$ 
increases, Fig.\ref{fig:traj} can be seen as a first, rough numerical clue of 
the emergence of typicality in the limit $N\gg1$.

\subsection{Maximal deviations from the average}

 In the next Section, using just simple mathematical methods of the probability 
theory, we will show that starting  from a  far-from-equilibrium initial 
condition (e.g. $|n_0-N/2|\gg \sqrt{N}$), $n_t$ exhibits an irreversible behavior in 
the limit $N\gg1$, namely it remains close to $\langle n_t\rangle$, which 
exponentially decays to its equilibrium value $N/2$.
More precisely, defining the quantity
\begin{equation}
\label{eq:delta}
 \delta_T = \sup_{0 \le t \le T} |n_t-\langle n_t \rangle|\,,
\end{equation}
i.e. the maximal deviation of $n_t$ from its average value along a trajectory of $T$ time-steps, we will show that
\begin{equation}
\label{eq:statementtoprove}
\mbox{Prob}\left(\frac{\delta_T}{N} <c_N \right)\ge 1-\zeta_N\,,
\end{equation}
where $T$ is $O(N)$, and both the constants $c_N$ and $\zeta_N$ tend to zero in the limit $N \to \infty$.

\begin{figure}[!h]
\centering
\includegraphics[clip=true,width=0.6\columnwidth,keepaspectratio]{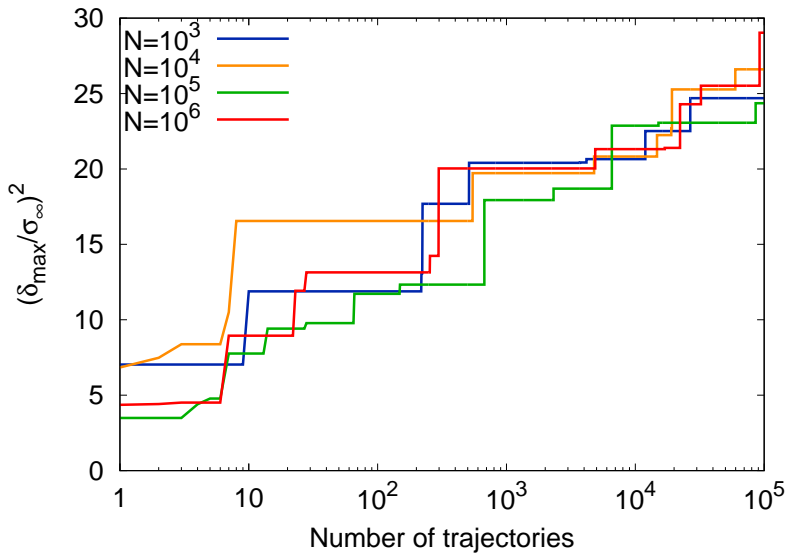}
\caption{
Scaling law between the maximal deviation $\delta_{max}$ from average as function of the 
number of trials $M$, for $N=10^3, 10^4, 10^5, 10^6$ 
and $T=2N$.  As in Fig.\ref{fig:traj}, we set $n_0=0.9N$ in all cases.}\label{fig:max}
\end{figure}

Before we exhibit a mathematical proof of relation~\eqref{eq:statementtoprove}, 
let us provide a numerical evidence of its validity. In Fig.~\ref{fig:max} we show
$\delta_{max}(M)$, defined as the maximal value of $\delta_T$ along $M$ 
different trajectories of length $T$, in function of $M$. The scaling of such a 
quantity with $M$ is particularly interesting, since the number of realizations 
we  need to observe a certain maximal deviation $\delta_{max}$ from the average 
is an indication of its probability.

Using Kac's lemma for recurrence times~\cite{kac47,kac57}, in \ref{Appendix:Kac} we derive the following scaling:
\begin{equation}
\label{eq:scaling}
{\delta_{max} \over \sqrt{N}} \sim \sqrt{ \ln M} \,,
\end{equation}
under the assumption that $\delta_T$, for large values, is distributed as a Gaussian
variable. Let us remark that detailed statistical features of extremal events can
be established in the framework of Gumbel’s theory~\cite{gumbel58}: however,
since we are only interested in the scaling law of $\delta_{max}$ vs $M$, we can avoid
the use of the complete theory by a direct application of Kac's lemma.
Let us remark that with a different distribution for $\delta_T$, still exponentially decaying to zero,
only minor changes occur in the result. We observe that Eq.\eqref{eq:scaling} is in fair
agreement with the numerical data 
presented in Fig.\ref{fig:max}.

Roughly speaking, Eq.~\eqref{eq:scaling} means that the maximal observed 
value of $\delta_T/\sqrt{N}$ grows very slowly with the number of trials $M$, 
implying that significative deviations from the average are extremely rare.

\section{A simple analytical result}
\label{Sec:4}

In order to understand how much a single realization of $n_t$ deviates from its 
average $\langle n_t\rangle$, i.e. how much a single trajectory is ``typical'' 
in the sense discussed in the previous sections, we focus on the 
probability: \begin{equation}
\mathcal{P}_{N,T} = \text{Prob}\Bigl( \left|n_t - \langle n_t \rangle \right| < \varepsilon_N \quad \forall t\,:\, 0 < t \le T\Bigr)
\end{equation}
where $\varepsilon_N$ is a constant (it only depends on $N$) and by $\langle n_t 
\rangle$ we mean  $\langle n_t |n_0\rangle$. Our goal is to prove 
that \begin{equation}
\label{eq:thesis}
 \mathcal{P}_{N,T} \rightarrow 1\,\quad \mbox{as} \quad N\rightarrow \infty\,,\quad \varepsilon_N/N \rightarrow 0, \quad T/N \rightarrow const.
\end{equation}
In other words, we aim at showing that a single realization is almost surely 
contained in a stripe $\pm\varepsilon_N$, at least up to times $\sim O(N)$, 
being $\varepsilon_N$ a quantity that grows slower than $N$ when $N$ tends to 
infinity.

First of all let us define the following sets:
\begin{equation}
\begin{aligned}
 \Omega_{\tau} &\equiv \{  t=k\tau\,:\, k=1,2, ...,  \lfloor T/\tau \rfloor  \} \\
\overline{\Omega}_{\tau} &\equiv \{t \in \mathbb{Z}^+ \, : \, t\le T, t \notin \Omega_{\tau} \} \,, 
 \end{aligned}
\end{equation}
where $\tau \in \mathbb{Z}$ and $\lfloor x \rfloor$ is the integer part of the 
real number $x$; $\Omega_{\tau}$ is the set of the equidistant 
discrete times $t\le T$ separated by $\tau$, while the remaining times of the trajectory form the set $\overline{\Omega}_{\tau}$.
Defining $\mathcal{A}_t$ as the event that $\delta n_t \equiv |n_t - \langle 
n_t \rangle|< \alpha_N$, where $\alpha_N<\varepsilon_N$ is a constant, we can write the following lower bound for $\mathcal{P}_{N,T}$:
\begin{equation}
 \mathcal{P}_{N,T}\ge \text{Prob}\Bigl[ \left( \delta n_t< \varepsilon_N, \forall t\in \overline{\Omega}_{\tau}\right)\; \text{and} \; \left( \mathcal{A}_t,  \forall t\in \Omega_{\tau}\right) \Bigr]
\end{equation}
that can be decomposed, using the definition of the conditional probability $P$, as
\begin{equation}
\mathcal{P}_{N,T} \ge  P\left(  \delta n_t< \varepsilon_N, \forall t\in \overline{\Omega}_{\tau} \; | \; \mathcal{A}_t, \forall t\in \Omega_{\tau} \right)\cdot \text{Prob}\left( \mathcal{A}_t, \forall t\in \Omega_{\tau}  \right)\,.
\end{equation}
Because of the Markovian character of the process, the above inequality can be written as:
\begin{equation}
\label{eq:twofactors}
\mathcal{P}_{N,T} \ge P\left( \delta n_t< \varepsilon_N, \forall t \in \overline{\Omega}_{\tau} \; | \; \mathcal{A}_t, \forall t\in \Omega_{\tau} \right) \prod_{t\in \Omega_{\tau}} P\left( \mathcal{A}_{t} | \mathcal{A}_{t-\tau} \right) .
\end{equation}
 We will see that, with a suitable choice of $\tau$, such decomposition allows 
us to prove our statement~\eqref{eq:thesis}. Now we need to study the two 
factors of the right hand side of Eq.~\eqref{eq:twofactors}: the first one can 
be estimated using the transition rules of the Markov chain~\eqref{eq:markov}, 
whereas for the second one we will apply the Chebyshev's inequality.

Since at every time step the value of $n_t$ can only increase (or decrease) by 1, and the same holds for
$\langle n_t \rangle$, we can write the following inequality:
\begin{equation}
\label{eq:inequality}
 \delta n_t \le \delta n_{k\tau} + 2 \tau\,,\quad \forall t \, : k\tau < t < (k+1) \tau\,,
\end{equation}
i.e. during $\tau$ time steps, the distance between a particular trajectory and the average cannot spread more than $2 \tau$. We could consider even stronger bounds on $\delta n_t$, but inequality~\eqref{eq:inequality} is enough to prove our result. 
In particular, if $\alpha_N+2 \tau < \varepsilon_N$ relation~\eqref{eq:inequality} implies 
\begin{equation}
\label{eq:one}
P\left( \delta n_t < \varepsilon_N, \forall t \in \overline{\Omega}_{\tau} \; | \; \mathcal{A}_t, \forall t\in \Omega_{\tau} \right) = 1\,.
\end{equation}
Let us notice that if we consider:
\begin{equation}
\label{eq:expon}
 \alpha_N=N^a\,,\quad \varepsilon_N=N^b\,,\quad \tau=\lfloor \varepsilon_N/3 \rfloor
\end{equation}
if $N$ is large enough,
equation~\eqref{eq:one} holds as soon as
\begin{equation}
\begin{aligned}
\label{eq:constraint}
 0<a<b<1\,.
\end{aligned}
\end{equation}

In order to evaluate the product in the left hand side of Eq.\eqref{eq:twofactors} we use a different strategy:
denoting with $\mathcal{A}_t^c$ the complementary event of $\mathcal{A}_t$, for every $t$ Chebyshev's inequality~\cite{gnedenko98} ensures that
\begin{equation}
\label{eq:Chebicevbound}
P\left(\mathcal{A}^c_{t} | \mathcal{A}_{0} \right) \leq \frac{\sigma^2_t}{\alpha^2_N}  \leq \frac{N^{1-2a}}{4}\,,
\end{equation}
where we have used the bound $\sigma_t^2 \le N/4$ (Eq.~\eqref{eq:variancebound_1} of \ref{Appendix:1}).
Noting that
\begin{equation}
\begin{aligned}
P\left( \mathcal{A}_t|\mathcal{A}_0  \right)& = P\left(\mathcal{A}_t|\mathcal{A}_{t-\tau}\right) P\left(\mathcal{A}_{t-\tau}|\mathcal{A}_0\right) + P\left(\mathcal{A}_t|\mathcal{A}^c_{t-\tau}\right) P\left(\mathcal{A}^c_{t-\tau}|\mathcal{A}_0\right) \\
&\leq P\left(\mathcal{A}_t|\mathcal{A}_{t-\tau}\right)+P\left(\mathcal{A}^c_{t-\tau}|\mathcal{A}_0\right)\,,
\end{aligned} 
\end{equation}
from \eqref{eq:Chebicevbound} we easily get 
 \begin{equation}
 \label{eq:minorcond}
\begin{aligned}
P\left(\mathcal{A}_t|\mathcal{A}_{t-\tau}\right) &\geq P\left(\mathcal{A}_t|\mathcal{A}_0\right) - P\left(\mathcal{A}^c_{t-\tau}|\mathcal{A}_0\right)\\
&=1-\left[  P\left(\mathcal{A}^c_{t-\tau}|\mathcal{A}_0\right) + P\left(\mathcal{A}^c_{t}|\mathcal{A}_0\right)  \right] \geq 1-  \frac{N^{1-2a}}{2}\,.
\end{aligned}
\end{equation} 
Finally, using relation~\eqref{eq:minorcond} to estimate the product in Eq.\eqref{eq:twofactors}, we find:
\begin{equation}
\label{eq:productestimate}
 \mathcal{P}_{N,T} \ge \prod_{t\in \Omega_{\tau}} P\left( \mathcal{A}_{t} | \mathcal{A}_{t-\tau} \right) \geq \left( 1 - \frac{N^{1-2a}}{2}\right)^{\lfloor T/\tau \rfloor}\,,
\end{equation}
where  $\lfloor T/\tau \rfloor$ is the cardinality of $\Omega_{\tau}$. If we choose $T$ to be proportional to $N$, we can always find a constant $\beta$, independent of $N$, such that
\begin{equation}
 \lfloor T/\tau \rfloor \le 2\beta N^{1-b}\,.
\end{equation}
It is easy to show that, if the constraint~\eqref{eq:constraint} holds and, in addition, we choose $a$ such that:
\begin{equation}
\label{eq:constraint2}
 a>1-\frac{b}{2}\,,
\end{equation}
 the right hand side of Eq.~\eqref{eq:productestimate} approaches to one as $\exp\left(-\beta N^{2-2a-b}\right)$ when $N\rightarrow \infty$ (note indeed that from inequalities~\eqref{eq:constraint} and~\eqref{eq:constraint2} one has $1-2a<0$). 
This completes the proof, since Eq.~\eqref{eq:expon} ensures that $\varepsilon_N/N \rightarrow 0$ in such limit.
\\
As an example, we have that the couple $b=0.8$ and $a=0.7$ satisfies the relations \eqref{eq:constraint} and \eqref{eq:constraint2}, giving
$\mathcal{P}_{N,T}\ge \exp\left(-\beta N^{-0.2}\right)$.



\section{Conclusions}
\label{Sec:5}
The time irreversibility is an experimental fact whose validity
must be accepted as a (quite obvious) empirical property of macroscopic systems.
On the other hand it is not easy at all  to build a coherent theory
that conciliates such macroscopic behavior with the reversible
nature of the laws at the microscopic level (i.e. Newton's equations).
The difficult point is to give a mathematical dignity to the great
visionary conjecture of Boltzmann, stating that in macroscopic systems 
an irreversible behavior occurs for the overwhelming majority of the
initial conditions.

One of the most important steps  in the ambitious program of formalizing the 
idea that  irreversibility is a typical property is due to 
Lanford: he  had been able to show the validity of the conjecture of Boltzmann 
for rarified gases in a suitable limit. Lanford proved his result only for a 
short time (of the order of one collision time); in addition, some authors claim that the 
use of the Lebesgue measure in the formulation of the idea of typicality is 
questionable.

In the present paper we  give an additional contribution, mainly at a 
pedagogical level, to support the Boltzmann's conjecture. For the Ehrenfest 
model we show a result which shares the same philosophy of the Lanford's work: 
in the limit $N \gg 1$, 
 a single trajectory of $n_t$ is very close to its mean value $\langle n_t \rangle$, with probability close to $1$.

 Due to the stochastic nature of our 
system, it is possible to obtain analytical results up to large times, and there 
is no ambiguity about the possible interpretations of typicality.

\appendix

\section{Derivation of Eqs.~\eqref{eq:conditionaln} and \eqref{eq:conditionalvar}}\label{Appendix:1}
Let us start with the derivation of Eq.\eqref{eq:conditionaln}.
Defining the variable $\Delta_t$ as a random quantity which takes values $\pm1$ with probabilities $1-n_t/N$ and $n_t/N$, respectively, we have the recurrence relation:
\begin{equation}
\label{eq:recurrence_rel}
n_{t+1} = n_t + \Delta_t.
\end{equation}
Taking the conditional average of the state $n_{t+1}$ for a given $n_t$, using Eq.\eqref{eq:recurrence_rel}, we get:
\begin{equation}
\label{eq:rec_rel2}
\langle n_{t+1} | n_t \rangle = n_t + \langle \Delta_t | n_t \rangle = n_t + \left(1-\frac{n_t}{N}\right) - \frac{n_t}{N} = \left(1- \frac{2}{N}\right) n_t +1
\end{equation}
where we have just applied the definition of $\Delta_t$.
Defining the variable $\delta_t=\langle n_t | n_{t-1}\rangle - N/2$, we can replace Eq.~\eqref{eq:rec_rel2} with a recursive relation for $\delta_t$:
\begin{equation}
\delta_{t+1} = \left( 1-\frac{2}{N}  \right) \delta_t\,.
\end{equation}
Fixing the initial state $n_0$, corresponding to some $\delta_0$, we finally get:
\begin{equation}
\delta_t = \left( 1-\frac{2}{N}\right)^t \delta_0\,,
\end{equation}
which, in terms of $\langle n_t|n_0\rangle$ and the initial state $n_0$, reads:
\begin{equation}
\label{eq:final_condavn}
\langle n_t | n_0\rangle = \frac{N}{2} + \left( 1- \frac{2}{N}  \right)^t \left( n_0 - \frac{N}{2}  \right)\,.
\end{equation}

Analogue calculations allows us to compute Eq.\eqref{eq:conditionalvar}, using the same strategy of Eq.\eqref{eq:rec_rel2}, we can obtain a relation for $\langle n_{t+1}^2 |n_t \rangle$:
\begin{equation}
\langle n_{t+1}^2 |n_t\rangle = 1+ \frac{N-4}{N} n^2_t + 2 n_t\,.
\end{equation}
Applying recursively this equation and using that $\langle n_0^2 |n_0\rangle=n_0^2$, we get:
\begin{equation}
\begin{aligned}
\langle n_t^2 | n_0 \rangle& =\frac{N (N+1)}{4}+ N \left( n_0 - \frac{N}{2}  \right)\left( 1-\frac{2}{N} \right)^t \\
&+ \left( n_0^2 + \frac{N}{4}\left( N-4n_0-1 \right)   \right) \left( 1-\frac{4}{N}  \right)^t\,.
\end{aligned}
\end{equation}
Using Eq.\eqref{eq:final_condavn} it is straightforward to derive the variance, conditioned to the initial value $n_0$:
\begin{equation}
\label{eq:final_appcondvar}
\begin{aligned}
 \sigma^2_t  &= \langle n^2_t|n_0\rangle - \langle n_t|n_0\rangle^2 \\
&= {N \over 4} + \left(1- {4 \over N} \right)^t\left[\left( n_0 - \frac{N}{2}\right)^2 - {N \over 4}\right]+\left(1- {2 \over N} \right)^{2t} \left( n_0 - \frac{N}{2}\right)^2 \,,
\end{aligned}
\end{equation}
which leads to Eq.\eqref{eq:conditionalvar}. As a remark, we note that Eq.\eqref{eq:final_appcondvar} is bounded by: 
\begin{equation}
\label{eq:variancebound_1}
\sigma^2_t \leq   \sigma^2_{\infty} = \frac{N}{4}\,,
\end{equation}
the stationary variance of the Ehrenfest model.

\section{Scaling law for $\delta_{max}(M)$}\label{Appendix:Kac}

Let us consider the Markov chain \eqref{eq:markov} starting with the initial 
condition $n=n_0$. For each trajectory, Eq.~\eqref{eq:delta} defines the largest 
deviation $\delta_T$ from the average (within a time $T$), and we have indicated
as $\delta_{max}(M)$ the maximal occurrence of this quantity in $M$ independent 
realizations. Our goal is to find an (approximate) relation between 
$\delta_{max}$ and $M$. We assume an asymptotic behavior of the probability density function,
$p(x)$, of the random variable $x=\delta_T/\sigma_{\infty}$ for a single trajectory.
 For instance, we can assume that such distribution decays with a 
Gaussian tail,
\begin{equation}
\label{eq:assuming_gauss}
 p(x) \simeq \alpha e^{-cx^2}\quad \mbox{for} \quad x \gg 1\,.
\end{equation}
The (average) number of independent attempts $M$ that are needed in order to observe a value 
of $x$ larger than $X$ is simply given by Kac's lemma~\cite{kac57}: 
\begin{equation}
 \langle M \rangle = \left(\int_{X}^{\infty} dx\,p(x)\right)^{-1}\,.
\end{equation}
With the assumption \eqref{eq:assuming_gauss}, in the limit $X\gg 1$, the 
above relation can be written as \begin{equation}
  \ln \langle M \rangle \simeq -\ln\left( const.\frac{e^{-c{X}^2}}{2X}\right)\simeq cX^2\,.
\end{equation}
Inserting $X=\delta_{max}/\sigma_{\infty}\sim\delta_{max}/\sqrt{N}$ into the 
above relation, we finally get the scaling law in Eq.~\eqref{eq:scaling}. It is 
easy to understand that the above argument for the logarithmic dependence of $X$ 
as a function of $\langle M \rangle$ is rather robust and does not depend on the 
details of the distribution: assuming an asymptotic shape $p(x)\sim 
\exp\left(-cx^{A}\right)$, one obtains
\begin{equation}
 \ln \langle M \rangle \sim x^A\,.
\end{equation}

\section*{Acknowledgments}
We thank M. Falcioni and R. Figari for a critical reading of the manuscript.

\medskip
\bibliography{biblio}

\end{document}